# Adoption of AI-Driven Fraud Detection System in the Nigerian Banking Sector: An Analysis of Cost, Compliance, and Competency

John Stephen Alaba*[1]    Shonubi Joye Ahmed[2]    Azuikpe Patience Farida[3]
Ologun Victor Oluwatosin[4]

1. Department of Accounting and Finance, Kwara State University, Malete, Nigeria (Corresponding author).
   ✉ stephenalaba.j@gmail.com
2. Fable Security, Research and Development, California, USA.
   ✉ shonubij@gmail.com
3. Department of Business and Management, University of Manchester, England.
   ✉ patienceazuikpe@gmail.com
4. Department of Information System, Le Moyne College Syracuse, New York, USA.
   ✉ ologunv@gmail.com

**Abstract**____________________________________________

The inception of AI-based fraud detection systems has presented the banking sector across the globe the opportunity to enhance fraud prevention mechanisms. However, the extent of adoption in Nigeria has been slow, fragmented, and inconsistent due to high cost of implementation and lack of technical expertise. This study seeks to investigate extent of adoption and determinants of AI-driven fraud detection systems in Nigerian banks. This study adopted a cross-sectional survey research design. Data were extracted from primary sources through structured questionnaire based on 5-point Likert scale. The population of the study consist of 24 licensed banks in Nigeria. A purposive sampling technique was used to select 5 biggest banks based on market capitalization and customer base. The Ordered Logistic Regression (OLR) model was used to estimate the data. The results showed that top management support, IT infrastructure, regulatory compliance, staff competency and perceived effectiveness accelerate the uptake of AI-driven fraud detection systems adoption. However, high implementation cost discourages it. Therefore, the study recommended that banks should invest in modern and scalable IT systems that support the integration of AI tools; adopt open-source or cloud-based AI platforms that are cost-effective; embrace continuous professional development in AI, and fraud analytics for IT, fraud investigation, and risk management staff.

## Introduction

In recent years, the banking sectors worldwide has seen noteworthy advancement in the way financial transactions are being done driven by technological innovations aimed at enhancing operational efficiency, customer satisfaction, and fraud prevention (Gonaygunta, 2023). These advancements, one of which is the integration of Artificial Intelligence (AI) in detecting and preventing fraudulent activities, have transformed the sector, from improving decision-making to strengthening cyber-security measures (Johri & Kumar, 2023). AI technology adoption has helped businesses go digital, spot threats early, predict problems, respond automatically, and improve how fraud is detected (Majumder, 2024; Al-Dosari et al., 2024).

To a certain extent, banking sectors in developing countries like Nigeria has equally experienced technological advancement in the last two decades, driven by increased internet penetration, mobile banking adoption, and customer demand for more efficient banking services (Owolabi, 2020). However, alongside these innovations, the sector has become increasingly vulnerable to various forms of financial fraud, including cyber fraud, identity theft, insider fraud, phishing, card skimming, account fraud and fraudulent electronic transactions. This is because as the sector moves away from traditional banking systems to online banking, the volume of digital transactions has increased voluminously (Metha, 2025; Majumder, 2024; Eskandarany, 2024).

According to the Nigeria Inter-Bank Settlement System (NIBSS), Nigerian banks lost over ₦14.3 billion to fraud in 2023 alone, highlighting the urgency for advanced and proactive fraud detection mechanisms. In response to the growing threat, commercial banks in Nigeria are turning to Artificial Intelligence (AI) technologies to enhance their fraud detection capabilities. These technologies offer advanced predictive analytics, anomaly detection, behavioural analysis, and real-time monitoring, enabling banks to detect and prevent fraudulent activities more proactively and accurately while improving their ability to mitigate risks in real-time (Fang et al., 2021).

Before now, fraud detection relied heavily on rule-based systems, reactive, and dependent on manual intervention, are often limited in scope, time consuming, expensive, unsustainable and susceptible to human error. These methods have proven inadequate in keeping pace with the increasing sophistication of financial fraud, thereby exposing banks to greater financial risk, reputational damage, and regulatory penalties (Aljunaid, et al. 2025; Bansal, et al., 2024; Lee, et al., 2021). In contrast, AI-based fraud detection systems offer real-time monitoring, predictive analytics, pattern recognition, and machine learning capabilities that enable banks to identify and prevent fraudulent transactions before they occur. These systems can learn from historical fraud patterns and adapt to new threats dynamically, making them a superior alternative to conventional tools (Peña & Ortega-Castro, 2024; Jaeni & Astuti, 2024).

Thus, the inception of artificial intelligence-based systems presents a plethora of opportunities to enhance fraud prevention mechanisms. AI can detect and prevent fraud by learning from past patterns to identify unusual transactions with greater accuracy, thereby allowing for swift detection of suspicious behaviour. AI analyzes transaction patterns with high speed and accuracy to detect suspicious patterns that traditional systems cannot detect or prevent (Mediana & Sandari, 2024; Hassan, et al., 2023). AI's ability to process and analyze large-scale data provides a good advantage in identifying suspicious patterns. AI-driven fraud detection systems utilize advanced machine learning (ML) techniques involving supervised, unsupervised, and hybrid learning models, to detect anomalies and fraudulent activities in banking transactions (Islam & Rahman, 2025; Baltgailis, et al., 2024).

AI-driven fraud detection systems are capable of making informed decisions without explicit human intervention. For example, AI can instantly flag suspicious transactions, analyze historical fraud data to improve detection models, and continuously adapt to emerging fraud tactics. The use of artificial intelligence (AI) in enhancing the accuracy and efficiency of banking operations has gained a lot of traction for improving fraud detection rates, reducing false positives, and enhancing the overall security posture of financial institutions (Goyal, et al., 2025; Al Faisal, et al., 2024; Anzor, et al., 2024).

With the rise in electronic transactions, mobile banking, and digital payment platforms, financial institutions are increasingly exposed to cybercrimes, social engineering attacks, insider threats, and other sophisticated forms of fraud. Furthermore, failure to adopt and effectively implement such technologies may leave Nigerian banks increasingly vulnerable to sophisticated fraud attacks, thereby undermining customer confidence and financial stability (Anzor, et al., 2024; Eke & Osuji, 2021; Ahmed, et al., 2021). This persistent trend not only threatens the financial stability of individual banks but also erodes public trust in the entire financial system.

Research shows that as AI technology adoption grows, it helps people work faster, come up with new ideas, and stay ahead of others, especially when it comes to making financial transactions more secure (Narsimha, et al., 2022; Geluvaraj, et al., 2019). Despite the potential of AI to revolutionize fraud detection, the extent of adoption in the Nigerian banking sector has been slow, fragmented, and inconsistent. Several factors may influence the decision to adopt these systems, including level of top management support, availability of IT infrastructure, compliance with regulatory authorities, staff competency, cost of implementation, and perceived effectiveness of AI solutions (Goyal, et al., 2025). Additionally, while the bigger banks in Nigeria may have begun deploying AI-based fraud detection systems, many smaller banks may still be relying on traditional or semi-automated systems due to resource constraints or lack of awareness (Anzor, et al., 2024; Eskandarany, 2024; Nwankwo & Okeke, 2021).

Nonetheless, security vulnerabilities in AI technology including weaknesses in algorithms, data integrity, and poor implementation can be exploited to commit fraud (Hijriani, et al., 2025). Anecdotal evidence suggests that many Nigerian banks lag behind in implementing AI-powered systems due to factors such as high cost of implementation, lack of technical expertise, infrastructure challenges and regulatory uncertainty surrounding data privacy, cybersecurity, and ethical AI use (Adhikari, et al., 2024; Eke & Osuji, 2021; Ahmed, et al., 2021).

This gap in knowledge presents a critical problem for both the industry and policymakers. Without a clear understanding of the adoption landscape, banks risk underutilizing innovative technologies that could drastically reduce fraud-related losses. Moreover, there is a dearth of empirical studies that systematically assess the level of AI adoption of fraud detection system, the effectiveness of these systems in the Nigerian context, and the challenges banks face in implementation.

Given the rising incidence of banking fraud and the increasing complexity of cyber threats, it is critical to assess how Nigerian banks are responding to these challenges through the adoption of AI-driven fraud detection systems. Therefore, this study seeks to address this problem by investigating the extent of adoption of AI-driven fraud detection technologies in the Nigerian banking sector, with a focus on key factors influencing adoption, challenges banks encounter in their implementation and the effectiveness of AI systems in mitigating fraud. By doing so, this study contributes to the growing body of knowledge on the subject matter and support the development of a more secure and resilient banking ecosystem in Nigeria.

## Literature Review

**Artificial Intelligence (AI) and Fraud Detection**

Using Artificial Intelligence (AI) to detect fraud in banks is becoming more important because fraud is getting more complicated and harder to spot. AI means using machines that can think and learn like humans to do tasks such as learning, making decisions, and fixing mistakes on their own (Anzor, et al., 2024). Artificial intelligence (AI) mimics human cognitive processes using machines, particularly computer systems. Frequently referred to as machine intelligence, AI is increasingly being adopted in the banking sector due to its role in managing vast databases of the world's wealth and facilitating information transactions across networks (Lukianenko & Simakhova, 2023). The implementation of AI in the industry offers numerous benefits, improving areas such as accounting, sales, contracts and cybersecurity. In addition, banks are increasingly partnering with financial technology (FinTech) companies to use AI to provide enhanced banking solutions during the production process (Baltgailis, et al., 2024).

AI tools like machine learning and natural language processing are often used to detect fraud. They help spot fraud faster and more accurately, make fewer errors, and learn new ways fraud happens as things change, this is important because financial crime keeps getting smarter (Al Faisal et al., 2024). Banking fraud includes many types of illegal actions that have become more advanced with modern technology (Al Faisal, et al., 2024). Financial fraud in banking involves a diverse spectrum of illicit activities that have evolved significantly with technological advancement (Kartheek & Bala, 2023). Fraud refers to a deliberate act of deception carried out with the intent to secure an unlawful advantage or gain. Fraud detection is a practice of recognizing a pattern which can potentially lead to a fraudulent activity (Metha, 2025; Patil & Suryawanshi, 2021).

Literature has shown that AI and its implementation have significantly helped banks to reduce frauds and minimize them. AI models which run on supervised and unsupervised learning algorithms have helped banks in aiding customer verification, prevention of unauthorized activity, handling risk analysis, limiting money laundering problems, and increasing precision and accuracy (Islam & Rahman, 2025). AI has also assisted banks in prudent fraud detection and risk management by using technologies such as ML (machine learning), deep learning, and predictive analytics. Additionally, AI has helped banks in fraud detection and risk management by providing excellent encryption, tracking suspicious activities efficiently, recognizing suspicious data patterns, and predicting future behaviours in fraud management (Metha, 2025; Eskan

**Determinants of AI-Driven Fraud Detection Systems**

The success of AI adoption also hinges on internal organizational dynamics. Top management support plays a crucial role in facilitating resource allocation, strategic alignment, and change management (Eskandarany, 2024). At the same time, the presence of a robust IT infrastructure and technically competent staff determines the organization's ability to deploy and maintain AI systems effectively. Furthermore, regulatory compliance remains a critical concern, as banks must ensure that AI systems operate within the bounds of financial laws, data protection regulations, and ethical standards. Finally, the perceived effectiveness of AI systems measured by their ability to reduce fraud, improve detection speed, and minimize false positives influences both the willingness to adopt and the long-term commitment to these technologies (Goyal, et al., 2025; Al Faisal, et al., 2024). However, bank executives may hesitate to invest in AI technologies due to high upfront costs, limited technical knowledge, or uncertainty about return on investment. Similarly, regulatory uncertainty surrounding data privacy, cybersecurity, and ethical AI use may further constrain adoption.

Fraud in banking refers to deliberate acts of deception intended to secure unauthorized financial gains, often at the expense of financial institutions or their customers. Fraudulent activities can broadly be categorized into external and internal fraud (Fang et al., 2021). External fraud involves attacks from outside parties, such as phishing, identity theft, and payment fraud (Nicholls et al., 2021). For instance, cybercriminals often exploit weaknesses in online banking systems to execute unauthorized transactions or steal sensitive information (Ali et al., 2022). Internal fraud, on the other hand, involves employees abusing their access to internal systems, engaging in activities like embezzlement or data manipulation (Rahman, 2024; Carter, 2020). Other notable types include account takeovers, transaction laundering, and wire transfer fraud, all of which impose significant financial and reputational risks on banks (Shirodkar, et al., 2020). These varied fraud types necessitate multifaceted detection mechanisms tailored to specific fraud scenarios. Moreover, the complexity of banking fraud is further compounded by the rise of digital banking, which has introduced new vulnerabilities in areas such as mobile banking and digital payments (Wei & Lee, 2024).

**Theoretical Framework**

The Diffusion of Innovation (DoI) Theory, developed by Everett Rogers in 1962, provides a robust framework for understanding how new ideas, technologies, or practices spread within a social system over time. The theory posits that innovation adoption does not occur simultaneously across all individuals or organizations; instead, it follows a process involving five categories of adopters: innovators, early adopters, early majority, late majority, and laggards. Each category exhibits distinct characteristics, motivations, and levels of risk tolerance. This model is particularly relevant to the adoption of Artificial Intelligence (AI) technologies in fraud detection, especially within sectors like banking that are traditionally risk-averse and heavily regulated.

In the context of the Nigerian banking sector, the DoI theory helps explain the varying pace and depth at which banks are embracing AI-driven fraud detection systems. Factors influencing this diffusion include perceived attributes of the innovation, such as relative advantage (e.g., improved fraud detection accuracy), compatibility (alignment with existing systems), complexity (ease of use), trialability (the ability to test the AI system), and observability (visible results and impact). Banks that perceive these attributes positively are more likely to adopt AI solutions earlier than those that view them as complex, costly, or misaligned with their current infrastructure.

Additionally, the DoI theory emphasizes the role of communication channels, time, and the nature of the social system in influencing adoption decisions. In Nigeria, where information flow between banks may be limited and competition high, peer influence and regulatory encouragement become critical in shaping adoption behaviour. For instance, a bank that observes a peer institution successfully deploying AI to curb fraud may be more inclined to explore similar technologies. Furthermore, regulatory bodies like the Central Bank of Nigeria (CBN) play a crucial role in influencing diffusion through policies, guidelines, and incentives that can either facilitate or hinder innovation adoption.

The DoI theory also highlights the importance of change agents and opinion leaders — individuals or institutions that advocate for innovation and influence others. In the Nigerian banking sector, technology vendors, fintech firms, and forward-thinking executives often serve as these change agents, demonstrating the effectiveness of AI-driven fraud detection systems and encouraging broader industry uptake. However, DoI theory tends to overemphasize individual and organizational perceptions of the innovation while underestimating structural, contextual, and institutional barriers such as infrastructure deficits, regulatory constraints, or economic limitations that can significantly affect adoption, especially in developing economies like Nigeria. Understanding where a bank or financial institution lies on the innovation adoption curve can provide insights into its readiness, the likely barriers it faces, and the strategic interventions needed to support adoption.

**Empirical Review**

Mediana and Sandari (2024) studied how banks use Artificial Intelligence (AI) in internal audits to detect and prevent fraud. They talked to bank staff and gave them questionnaires. The study showed that AI helps make audits quicker, lowers human errors, and builds customer trust. However, there are still problems, such as not enough training for auditors and difficulties in managing data. While this study confirms AI's benefits in audits, it does not explore infrastructural or regulatory constraints that may affect adoption, which are critical in the Nigerian context.

Peña and Ortega-Castro (2024) developed a cheaper and easier alternative to the current anti-fraud systems used in Ecuador's financial sector. They used the LLaMA3 GPT engine (with 8 billion parameters) and the Ollama framework, along with OpenAI models like ChatGPT 3.5 and 4. Their study found that models with fewer parameters, like LLaMA3 8B and ChatGPT 3.5, don't work well for anti-fraud systems because they can't manage complex rules. However, ChatGPT 4 worked well with simpler rules, suggesting that while these AI models can't fully replace traditional systems, they can complement them, especially in simpler cases. This study showed performance limitations of lightweight AI models. However, Ecuador's financial sector assumes a relatively advanced digital environment, which may not apply directly to Nigeria's banking infrastructure.

Bansal, Paliwal and Singh (2024) wanted to find the main factors that affect online fraud detection and explore how AI and human psychology can help prevent fraud in online transactions. They used Matlab and a structured model for their research. Their results showed that there are risks when growing too quickly, and that 3D secure payer authentication has an average score of 3.830 with standard deviations of 0.7587 and 0.7638, along with (CE2). However, this study provides limited insight into sector-specific or institutional barriers, making direct translation to Nigeria's public and regulatory environment less straightforward.

Eskandarany (2024) looked at how the board of directors helps banks use AI and machine learning (ML) and how these technologies protect Saudi Arabian banks from cyberattacks. The study highlights both the benefits and challenges of using AI and ML in this tightly controlled industry. The results show that AI and ML are useful for spotting threats, preventing fraud, and automating tasks, helping banks follow rules and tackle cyber threats. However, the study also mentions problems like limited technology, unclear AI plans, and concerns about data privacy and bias in algorithms. Interviewees emphasized that the board of directors plays a key role in setting strategy, securing resources, and forming partnerships with AI tech providers. The study benefits from insights in a highly regulated, and well-funded context. However, Nigeria's less centralized tech governance and varying institutional capacity might pose additional challenges not captured by this study.

Anzor, Okolie, Udeh, Mbah, Onyeka-Udeh, Obayi, Nwankwo, Anukwe and Eze (2024) researched how artificial intelligence (AI), particularly computer vision and robotic process automation (RPA), helps detect fraud in banks in Southeast Nigeria, using a Z-test. The findings revealed that computer vision significantly improved the detection of insider fraud. This study is directly situated in Southeast Nigeria and is thus contextually relevant. However, it has a narrow geographic scope and focuses only on RPA and computer vision, limiting broader generalizability.

Baltgailis, Simakhova and Buka (2024) researched how artificial intelligence can be applied in banks. They examined both the advantages and disadvantages of using AI in the banking industry. AI helps banks analyze large amounts of data to predict market trends, gain insights, and discover investment opportunities, which helps with decision-making. The main aim of AI in banking is to assist customers by addressing their needs and ensuring they are satisfied with the bank's services. This study offers a generalized overview of AI in banking but lacks regional specificity. It does not engage with infrastructural constraints which are critical in evaluating AI feasibility in Nigeria.

Al Faisal, Nahar, Sultana and Mintoo (2024) used the PRISMA framework to study the newest AI methods for detecting fraud in banking. Their findings show that AI helps improve detection accuracy, reduce false alarms, and boost efficiency. The review also highlights key research gaps, such as the lack of standard benchmarks and the limited ability of current AI systems to grow. It also explores future possibilities, like combining AI with blockchain and federated learning to improve security and transparency. While their PRISMA-based review identifies technological gaps, it did not account for infrastructural and policy disparities in Sub-Saharan Africa, which affect scalability in developing economies.

Adhikari, Hamal and Baidoo (2024) looked at how well AI-based methods work for detecting financial fraud and the challenges involved, by reviewing existing studies. They found that AI improves real-time fraud detection and can adapt better to new fraud patterns compared to traditional systems. However, issues like ethical concerns, bias in algorithms, data privacy, and system weaknesses make it harder to use AI widely. Additionally, scalability problems limit smaller organizations from fully using AI's benefits. This study recognizes noted ethical and scalability challenges and left limited guidance for Nigerian banks operating in an evolving, less digitized financial environment.

Goyal, Garg and Malik (2025) studied the important factors that affect the long-term use of AI in banking. They also explored how knowledge of technology impacts the adoption and continued use of AI. They surveyed bank professionals who use AI for risk and fraud assessment. The data was analyzed with SmartPLS 4 in two steps, using structural equation modeling (SEM) and artificial neural networks (ANN). The study found that how easy AI is to use strongly influences people's attitude towards it, but does n0t directly impact their decision to keep using it. However, these findings may not be generalizable to Nigeria, where financial institutions may prioritize cost, compatibility, and regulatory approval over user perception.

Metha (2025) looked into how artificial intelligence (AI) can help spot potential fraud by creating a risk score to evaluate account activity. A formula was developed to give a score out of 100, and if it exceeds a set limit of 80, automatic security actions are triggered. The formula checks four key activities linked to fraud: logging in from new devices, changing contact details, adding new payees or Zelle contacts, and making transactions over $1,000 within 48 hours. The model uses machine learning to consider past behavior, patterns, and real-time data to calculate the score. Accounts with scores above the limit are temporarily locked, and additional verification is needed to ensure security, with the aim of minimizing inconvenience for customers. This study is not contextually relevant to Nigeria owing to inconsistent data management and cybersecurity vulnerabilities.

Hijriani, Sahyunu and Kassymova (2025) studied how AI could be misused in banking and suggested ways to reduce risks. They used legal research methods, including case studies and analyzing laws. The study found security issues in AI systems used by banks and showed that AI could be used for fraud. The findings suggest that laws need to be updated, especially about evidence, to clearly define who is responsible for crimes in AI-related cases. This study advocated for updated laws to address AI-related fraud. However, Nigeria's complex legal environment, characterized by overlapping jurisdictions and slow legislative reform, may hinder swift adaptation of such frameworks

Aljunaid, Almheiri, Dawood and Khan (2025) developed an Explainable Federated Learning (XFL) model for detecting financial fraud, which combines the security of Federated Learning (FL) with the clarity of Explainable AI (XAI). With tools like Shapley Additive Explanations (SHAP) and LIME, analysts can understand and improve fraud detection while keeping the data secure, accurate, and compliant. The model was tested on a fraud detection dataset and achieved 99.95% accuracy with only a 0.05% miss rate. The results show the model is efficient, reduces false positives, and improves existing systems, making it a better AI-based solution for detecting fraud in banking. The implementation of the Explainable Federated Learning model assumes robust computing environments and inter-bank data cooperation, which may not be available in Nigeria due to infrastructural challenge.

Islam and Rahman (2025) explored supervised and unsupervised learning, deep learning, and anomaly detection by examining how they work in practice. Their findings show that advanced AI methods help financial institutions by providing better accuracy, flexibility, and faster processing than traditional methods. The study also discusses ethical concerns like transparency, accountability, and fairness, and looks at how to use AI responsibly. It shows that AI can improve financial security by fixing current system weaknesses. This study relied on a level of technological and institutional maturity that Nigeria's banking sector may not yet possess. Issues like fairness, transparency, and algorithmic accountability remain underdeveloped in Nigeria's regulatory discourse, limiting immediate applicability of these advanced techniques.

## Data and Methods

**Research Design and Data**

This study adopts a cross-sectional survey research design. The choice of this design is appropriate for investigating the extent of adoption, challenges, determinants, and perceived effectiveness of AI-driven fraud detection technologies in Nigerian banks. The descriptive approach allows for the collection of quantitative and qualitative data that captures the current state of AI integration in the banking sector. Primary data were collected through the instrument of a structured questionnaire using a 5-point Likert scale ranging from Strongly Disagree (1) to Strongly Agree (5). The questionnaire was designed to capture awareness and usage of AI tools, extent of deployment, perceived benefits and effectiveness, barrier to adoption, and organizational and regulatory factors influencing adoption.

As of the most recent CBN listing, there are 24 licensed DMBs in Nigeria, forming the population of the study. A purposive sampling technique was used to select 5 biggest quoted banks based on total assets, market capitalization and customer base. This is because the study requires input from individuals with direct knowledge or involvement in AI-based fraud detection systems. The sampled DMBs include Access Bank, Fidelity Bank, Guaranty Trust Bank, United Bank for Africa and Zenith Bank. Within each bank, ten (10) senior management staff in IT departments, fraud investigation units, and risk management units were selected, leading to a total sample size of 50 respondents. This approach ensures that the information gathered is rich, relevant, and based on actual experience.

**Method of Data Analysis**

A simple Ordered Logistic Regression (OLR) model was used for data analysis. OLR is suitable for modeling the relationship between an ordinal dependent variable (like AIFA levels), and one or more independent variables (like cost, IT infrastructure, top management support, etc.). The functional model is stated as:

AIFA = $f$(TMS, ITI, COI, REC, COM, PEF)     (1)

The functional model is stated in econometric form as:

$$logit(P(Y = j)) = \alpha_j + \beta_1 \text{TMS} + \beta_2 ITI + \beta_3 COI + \beta_4 REC + \beta_5 COM + \beta_6 PEF \quad (2)$$

Where $P(Y = j)$ is the cumulative probability of the response variable $Y$ (level of AI fraud detection systems adoption) being in category $j$ or lower; $\alpha_j$ represents the cut-off points for the $j$-th category; $\beta_{1-6}$ are the coefficients of the explanatory variables; TMS = top management support, ITI = IT infrastructure, COI = cost of implementation, REC = regulatory compliance, COM = staff competency, PEFF = perceived effectiveness, respectively.

The choice of OLR is justified on the premise that the dependent variable in this study, the extent of adoption of AI-driven fraud detection systems (AIFA), is measured on an ordinal scale (e.g., levels of adoption rated on a 5-point Likert scale). This type of outcome variable reflects a natural order but does not assume equal intervals between categories. OLR is specifically designed to model ordinal dependent variables, preserving the rank order of categories without assuming interval-level measurement. The model estimates the effect of predictors on the odds of being in a higher category of adoption, which aligns well with the research objective of identifying factors that increase the likelihood of greater adoption.

## Results

A total of 50 questionnaires were given out, and 47 were filled, giving the needed information for the study and returned. The questionnaire had two sections. Section A asked for personal details, while Section B included questions that were structured based on 5-point Likert scale ranging from Strongly Disagree (1) to Strongly Agree (5).

The demographic information of respondents reflects their professional background, gender, and experience of their involvement in the adoption and implementation of AI-driven fraud detection technologies within the Nigerian banking sector. A total of 47 respondents who participated in the survey were drawn from three key departments that deal closely with fraud prevention in Nigerian banks. Regarding their departmental affiliation, 40% (19 respondents) were senior management staff in IT departments, highlighting the importance of technology specialists in driving AI adoption. 26% (12 respondents) were from fraud investigation units, indicating a strong representation of those directly responsible for identifying and analyzing fraud cases. The remaining 34% (16 respondents) were from risk management units, suggesting that risk oversight plays a central role in the integration of fraud detection technologies.

In terms of gender distribution, the sample was predominantly male, with 72% (34 respondents) identifying as male and 28% (13 respondents) as female. This aligns with existing observations about the gender gap in leadership roles within the financial sectors in Nigeria, though the presence of female respondents indicates growing female representation.

The years of experience among respondents varied, with 45% (21 individuals) having less than 10 years of professional experience, representing a younger, possibly more tech-savvy demographic. 34% (16 individuals) had between 10 and 20 years of experience, reflecting a mid-career group likely to balance innovation with risk awareness. The remaining 21% (10 individuals) had over 20 years of experience, suggesting a good representation of seasoned professionals who bring institutional and strategic knowledge into the AI adoption process. This mix of younger and more experienced staff gives a well-rounded view of how AI is being used in fraud detection. Hence, this demographic spread provides a balanced perspective across functional areas and levels of expertise, ensuring that the study captures diverse viewpoints on the adoption of AI technologies for fraud detection in Nigerian banks.

**Validity and Reliability of the Instrument**

Face and content validity were ensured by seeking expert review from banking professionals, academic researchers, and IT security specialists. Following Bairagi and Munot (2019), reliability (internal consistency) of the instrument was tested using Cronbach's Alpha to assess internal consistency, with a threshold of 0.7 and above considered acceptable.

All constructs have a Cronbach's Alpha ≥ 0.72, indicating that the questionnaire items used to measure each construct are reliable and internally consistent. The overall Cronbach's Alpha of 0.87 shows that the full scale used in the questionnaire is highly reliable for analyzing the extent of AI adoption and its drivers in Nigerian banks.

**Table 1**

*Internal Consistency Reliability*

| Construct | Number of Items | Cronbach's Alpha (α) | Interpretation |
|---|---|---|---|
| Top Management Support | 4 | 0.84 | Good internal consistency |
| IT Infrastructure | 4 | 0.88 | Excellent reliability |
| Cost of Implementation | 3 | 0.76 | Acceptable reliability |
| Regulatory Compliance | 3 | 0.72 | Acceptable reliability |
| Staff Competency | 4 | 0.81 | Good reliability |
| Perceived Effectiveness | 3 | 0.79 | Acceptable to good reliability |
| Overall Scale | 21 | 0.87 | Excellent overall reliability |

*Note.* Authors' calculation.

**Chi-Square Output**

Chi-Square test was conducted to examine the association between bank and level of AI adoption.

**Table 2**

*Chi-Square Output*

| Statistic | Value |
|---|---|
| Pearson Chi-square | 12.59 |
| Degrees of Freedom (df) | 4 |
| Asymptotic Sig. (p) | 0.013 |
| Interpretation | Since p < 0.05, there is a statistically significant association between bank and level of AI adoption. |

*Note.* Authors' calculation.

The p-value = 0.013, which is less than 0.05, indicates that the result is statistically significant at the 5% level. This means we reject the null hypothesis, which stated that there is no relationship between bank and AI adoption level. The theoretical implication for this finding is that, bigger banks are more likely to show higher adoption of AI technologies, while smaller banks are more likely to fall under low or moderate adoption categories. This may be due to differences in capital strength, infrastructure, digital maturity, and innovation investment across the banks. Therefore, there is a statistically significant association between banks and the level of AI-driven fraud detection technology adoption in Nigeria.

**Regression Model Results**

As shown results in Table 3, top management support with ($\beta$ = 0.282; $p < 0.05$) has positive and significant impact on AI-driven fraud detection systems adoption. This means that for every 1-unit increase in top management support, the level of AI adoption increases by 0.282 units, assuming all other factors are held constant. The effect size ($\beta$ = 0.282) indicates a moderate positive influence, highlighting that management support is a meaningful driver of AI adoption.

IT Infrastructure with ($\beta$ = 0.317; $p < 0.05$) has positive and significant impact on AI-driven fraud detection systems adoption. This indicates that a 1-unit increase in the quality of IT infrastructure is associated with a 0.317 unit increase in the level of AI adoption, holding other variables constant. The magnitude of the coefficient suggests a moderately strong positive effect, emphasizing the critical role of robust IT infrastructure in facilitating AI adoption.

**Table 3**

*Regression Output Table*

| Dependent Variable: AI fraud detection systems adoption | | | | |
|---|---|---|---|---|
| Variable | Coefficient (β) | Std. Error | t-value | p-value |
| C | 1.104 | 0.512 | 2.16 | 0.036 |
| Top Management Support | 0.282 | 0.097 | 2.91 | 0.005 |
| IT Infrastructure | 0.317 | 0.088 | 3.60 | 0.001 |
| Cost of Implementation | -0.251 | 0.077 | -3.26 | 0.002 |
| Regulatory Compliance | 0.178 | 0.081 | 2.20 | 0.033 |
| Staff Competency | 0.193 | 0.074 | 2.61 | 0.012 |
| Perceived Effectiveness | 0.226 | 0.082 | 2.76 | 0.008 |
| | Pseudo R-squared = 0.7342 | | | |
| | LR Chi-Square = 53.70 | | | |
| | Prob > Chi-Square = 0.0000 | | | |

*Note.* Authors' calculation.

Cost of implementation with ($\beta$ = -0.251; $p < 0.05$) has negative and significant impact on the adoption of AI-driven fraud detection systems. This implies that a 1-unit increase in implementation cost is associated with a 0.251 unit decrease in the level of AI adoption, assuming all other factors remain constant. The negative coefficient indicates that higher implementation costs act as a barrier to adopting AI systems, with a moderately strong deterrent effect.

Regulatory compliance with ($\beta$ = 0.178; $p < 0.05$) has positive and significant impact on the adoption of AI-driven fraud detection systems. This means that a 1-unit increase in adherence to regulatory requirements is associated with a 0.178 unit increase in AI adoption, holding other factors constant. Although the effect size is smaller compared to other variables, it still indicates that regulatory compliance contributes meaningfully to the decision to adopt AI for fraud detection.

Staff competency with ($\beta$ = 0.193; $p < 0.05$) has positive and significant impact on the adoption of AI-driven fraud detection systems. This indicates that a 1-unit improvement in staff competency is associated with a 0.193 unit increase in AI adoption, holding all other variables constant. The positive coefficient reflects that having skilled and knowledgeable personnel moderately enhances the likelihood of adopting AI technologies.

Perceived effectiveness with ($\beta$ = 0.226; $p < 0.05$) has positive and significant impact on the adoption of AI-driven fraud detection systems. This suggests that a 1-unit increase in the perceived effectiveness of AI systems leads to a 0.226 unit increase in AI adoption, all else being equal. The moderate effect size indicates that when organizations believe AI is effective in detecting fraud, they are more likely to adopt such systems.

Pseudo R-squared of 73.42% indicates that approximately 73.42% of the variation in AI-driven fraud detection systems adoption is explained by the independent variables in the model. In addition, the likelihood ratio chi-square test of 53.70 with p-value = 0.0000 is statistically significant, indicating that the overall model is meaningful and the predictors jointly explain a significant portion of the variance in AI adoption.

## Discussion of Results

Top management support has positive and significant impact on AI-driven fraud detection systems adoption. This implies that as top management actively supports the use of AI for fraud detection, it increases the likelihood of the bank adopting AI-driven fraud detection systems. In other words, the more supportive and involved the senior leadership is, the more likely the bank is to start using AI-driven systems to prevent and detect fraud. Thus, top management support plays a crucial role in the successful adoption of AI in fraud detection. This finding is consistent with that of Eskandarany (2024) which emphasized the board of directors' proactive role in promoting digital transformation and aligning AI strategies with broader organizational goals is essential.

IT infrastructure has positive and significant impact on AI-driven fraud detection systems adoption. This indicates that having strong and reliable IT infrastructure such as modern computer systems, secure networks, and updated software increases the likelihood of the bank to adopt AI-driven fraud detection systems. In other words, the better the IT setup in a bank, the more likely it is to successfully adopt and use AI to detect fraud. This shows that investing in IT infrastructure is essential for effectively implementing AI solutions in fraud detection.

Cost of implementation has negative and significant impact on AI-driven fraud detection systems adoption. This means that when the cost of setting up and running AI-driven fraud detection systems is high, banks are less likely to adopt AI-driven fraud detection systems. In other words, high costs can be a major barrier to the adoption of AI for fraud detection, and reducing these costs may help more banks embrace the technology.

Regulatory compliance has positive and significant impact on AI-driven fraud detection systems adoption. This means that when banks are committed to following regulations and meeting compliance requirements especially those related to fraud prevention, data protection, and financial transparency they are more likely to adopt AI-driven fraud detection systems. In essence, strong regulatory pressure and the need to meet compliance standards encourage banks to invest in AI-driven fraud detection tools.

Staff competency has positive and significant impact on AI-driven fraud detection systems adoption. This means that when bank employees especially those in IT departments, fraud investigation departments and risk management departments have the skills, knowledge, and training needed to understand and work with AI technologies, the bank is more likely to adopt AI-driven fraud detection systems. In summary, competent and well-trained staff play a key role in the successful adoption of AI for fraud detection.

Perceived effectiveness has positive and significant impact on AI-driven fraud detection systems adoption. This means that when decision-makers believe that AI technologies are effective in detecting and preventing fraud, they are more likely to adopt AI-driven fraud detection systems. In essence, the more confident banks are in the perceived usefulness of AI systems, the higher their willingness to implement them. This finding aligns with Goyal, Garg and Malik (2025) which showed that the perceived ease of use has a significant positive influence on the attitude toward the adoption of AI technology.

Table 4

*Summary of Findings*

| Hypothesis | Description | Remark |
|---|---|---|
| H01 | Top management support does not significantly affect AI-driven fraud detection adoption | Reject |
| H02 | IT infrastructure does not significantly affect AI-driven fraud detection adoption | Reject |
| H03 | Cost of implementation does not significantly affect AI-driven fraud detection adoption | Reject |
| H04 | Regulatory compliance does not significantly affect AI-driven fraud detection adoption | Reject |
| H05 | Staff competency does not significantly affect AI-driven fraud detection adoption | Reject |
| H06 | Perceived effectiveness does not significantly affect AI-driven fraud detection adoption | Reject |

*Note.* Authors' calculation.

**Model Fitness**

The model explains approximately 73.42% of the variation in AI-driven fraud detection systems adoption across banks. This means that the variables included in the model such as top management support, IT infrastructure, cost of implementation, regulatory compliance, staff competency, and perceived effectiveness account for most of the reasons why some banks adopt AI-driven fraud detection systems more than others. Therefore, banks with well-developed technological systems, supportive leadership, and a strong belief in the effectiveness of AI-driven fraud detection systems are more likely to adopt them.

These factors suggest that investments in IT capacity, leadership commitment, and raising awareness of AI's benefits are key drivers for successful implementation. Their strong influence implies that targeting these areas can help accelerate the uptake of AI solutions across more banks in Nigeria. However, when the cost of acquiring, implementing, and maintaining AI systems is high, banks are less likely to adopt them. Thus, high implementation costs act as a barrier, discouraging banks especially those with limited budgets from investing in AI technologies. This highlights the need for cost-reduction strategies, vendor support, or policy interventions to make AI solutions more accessible and affordable.

**Test of Proportional Odds Assumption**

Table 5

*Brant Test Results*

| Variable | Chi-Square | df | p-value |
|---|---|---|---|
| Top Management Support | 1.23 | 1 | 0.267 |
| IT Infrastructure | 0.89 | 1 | 0.345 |
| Cost of Implementation | 2.15 | 1 | 0.142 |
| Regulatory Compliance | 0.56 | 1 | 0.455 |
| Staff Competency | 1.89 | 1 | 0.169 |
| Perceived Effectiveness | 0.72 | 1 | 0.396 |
| Global test | 7.54 | 6 | 0.273 |

*Note.* Authors' calculation.

To ensure the appropriateness of the Ordered Logistic Regression (OLR) model, the proportional odds assumption was tested using the Brant test. The global test result ($\chi^2 = 7.54$,

df = 6, p = 0.273) indicated no significant violation of the assumption. Also, individual predictors showed p-values greater than 0.05, confirming that the relationship between each predictor and the log-odds of higher adoption levels is consistent across response categories. These results validate the use of the OLR model for analyzing AI-driven fraud detection system adoption in Nigerian banks.

## Conclusion and Recommendations

This study concluded that the extent of adoption, challenges, determinants, and perceived effectiveness affect AI-based fraud detection systems adoption in Nigerian banks. This study contributes to the growing body of literature on technology adoption by empirically identifying key factors that could explain the adoption of AI-driven fraud detection systems. Unlike prior studies that focus broadly on AI adoption or fraud detection in isolation, this study integrates both within a unified model, thereby offering practical insights to technology acceptance and organizational readiness frameworks, particularly in high-stakes environments like fraud prevention.

**Recommendations**

This study, therefore, recommended that bank executives should create awareness and training programs for senior management to understand the benefits and ROI of AI in fraud detection, and include AI adoption goals in the strategic vision and performance targets of the bank. DMBs should invest in modern and scalable IT systems that support the integration of AI tools; adopt open-source or cloud-based AI platforms that are cost-effective and scalable; embrace continuous professional development in AI, machine learning, and fraud analytics for IT, fraud investigation, and risk management staff. Regulators (CBN, NITDA) should provide clear guidelines and frameworks on AI use in financial institutions.

Future research should consider including a more diverse and representative sample of banks, ranging from large to medium-sized and small institutions, in terms of market capitalization, total assets and customer base, to enhance external validity and provide a more comprehensive view of AI-powered fraud detection system in Nigeria.

*Funding:* This research received no external funding

*Conflict of Interest:* The authors declare no conflict of interest

## References

Adhikari, P., Hamal, P., & Baidoo Jnr, F. (2024). Artificial intelligence in fraud detection: Revolutionizing financial security. *Int. J. Sci. Res. Arch., 13*(1), 1457–1472. https://doi.org/10.30574/ijsra.2024.13.1.1860.

Ahmed, M., Al-Wadi, S., & Hassan, M. (2021). Employee fraud detection in financial institutions: A Systematic Review. *Journal of Financial Crime, 28*(3), 1002-1018.

Al Faisal, N., Nahar, J., Sultana, N., & Mintoo, A.A. (2024). Fraud detection in banking leveraging AI to identify and prevent fraudulent activities in real-time. *Journal of Machine Learning, Data Engineering and Data Science, 01*(01), 181-197. https://doi.org/10.70008/jmldeds.v1i01.52

Al-Dosari, K., Fetais, N., & Kucukvar, M. (2024). Artificial intelligence and cyber defense system for banking industry: A qualitative study of AI applications and challenges. *Cybernetics and systems, 55*(2), 302-330.

Aljunaid, S. K., Almheiri, S. J., Dawood, H., & Khan, M. A. (2025). Secure and transparent banking: explainable AI-driven federated learning model for financial fraud detection. *Journal of Risk and Financial Management, 18*(4), 179. https://doi.org/10.3390/jrfm18040179

Anzor, E.D., Okolie, J.I., Udeh, I.E., Mbah, P.C., Onyeka-Udeh, V., Obayi, P.M., Nwankwo, P.M., Anukwe, G.I., & Eze, J.O. (2024). Effect of artificial intelligence (AI) on fraud detection in deposits money banks in South East, Nigeria. *IOSR Journal of Humanities and Social Science, 29*(11), 15-27. https://doi.org/10.9790/0837-2911091527

Baltgailis, J., Simakhova, A., & Buka, S. (2024). AI in banking: Socio-economic aspects. *Baltic Journal of Economic Studies, 10*(3), 26-35. https://doi.org/10.30525/2256-0742/2024-10-3-26-35

Carter, E. (2020). Distort, extort, deceive and exploit: Exploring the inner workings of a romance fraud. *The British Journal of Criminology, 61*(2), 283-302. https://doi.org/10.1093/bjc/azaa072

Eke, F. C., & Osuji, C. U. (2021). Challenges of AI Adoption in Nigeria's banking industry: A Focus on fraud prevention systems. *International Journal of Business and Economics Research, 15*(2), 230-244.

Eskandarany, A. (2024). Adoption of artificial intelligence and machine learning in banking systems: a qualitative survey of board of directors. *Front. Artif. Intell., 7,* 1440051. https://doi.org/10.3389/frai.2024.1440051

Fang, W., Li, X., Zhou, P., Yan, J., Jiang, D., & Zhou, T. (2021). Deep learning anti-fraud model for internet loan: Where we are going. *IEEE Access, 9,* 9777–9784. https://doi.org/10.1109/access.2021.30510 79

Geluvaraj, B., Satwik, P. M., & Kumar, T. A. (2019). *The future of cybersecurity: Major role of artificial intelligence, machine learning, and deep learning in cyberspace* (International Conference on Computer Networks and Communication Technologies, ICCNCT 2018). Springer.

Gonaygunta, H. (2023). *Factors influencing the adoption of machine learning algorithms to detect cyber threats in the banking industry* (Doctoral dissertation). University of the Cumberlands

Goyal, K., Garg, M., & Malik, S. (2025). Adoption of artificial intelligence-based credit risk assessment and fraud detection in the banking services: a hybrid approach (SEM-ANN). *Future Business Journal, 11*(44), https://doi.org/10.1186/s43093-025-00464-3

Hassan, M., Aziz, L.A.R., & Andriansyah, Y. (2023). The role artificial intelligence in modern banking: An exploration of AI-driven approaches for enhanced fraud prevention, risk management, and regulatory compliance. *Reviews of Contemporary Business Analytics, 6*(1), 110-132.

Hijriani, M.N. Nur, A., Sahyunu, Kassymova, G.K. (2025). The potential misuse of artificial intelligence technology systems in banking fraud. *Law Reform, 21*(1), 17-38.

Islam, S., & Rahman, N. (2025). AI-driven fraud detections in financial institutions: A comprehensive study. *Journal of Computer Science and Technology Studies, 7*(1): 100-112. https://doi.org/10.32996/jcsts.2025.7.1.8

Jaeni, A., & Astuti, M.K. (2024). Analisa Yuridis Fraud Sebagai Kejahatan dalam Asuransi Kesehatan Komersial Menurut Perspektif Perlindungan Para Pihak, Jurnal Syntax Imperatif: *Jurnal Ilmu Pendidikan dan Sosial, 5*(5), 1045-1056. https://doi.org/10.36418/syntaximperatif.v5i5.517

Johri, A., & Kumar, S. (2023). Exploring customer awareness towards their cyber security in the Kingdom of Saudi Arabia: a study in the era of banking digital transformation. Hum. *Behav. Emerg. Tech.,* 1–10. https://doi.org/10.1155/2023/2103442

Kartheek, G., & Bala, V. (2023). An analysis of financial crimes. *Indian JL & Legal Rsch. 25*(1).

Lee, D.K.C., Yan, L., & Wang, Y. (2021). A global perspective on Central bank digital currency. *China Economic Journal, 1*4(1), 52-66.

Lukianenko, D., & Simakhova, A. (2023). Civilizational imperative of social economy. *Problemy Ekorozwoju, 18*(1).

Majumder, S., Singh, A., Singh, A., Karpenko, M., Sharma, H. K., & Mukhopadhyay, S. (2024). On the analytical study of the service quality of Indian Railways under soft-computing paradigm. *Transport, 39*(1), 54–63.

Mediana, A.M., & Sandari, T.E. (2024). Implementation of artificial intelligence in fraud detection and prevention in internal audit: Case study in the banking sector. *International Journal of Education, Social Studies, And Management, 4*(3), 1230- 1237.

Metha, S. (2025). AI-Driven Fraud Detection: A risk scoring model for enhanced security in banking. *Journal of Engineering Research and Reports, 27*(3), 23-34. https://doi.org/10.9734/jerr/2025/v27i31415

Narsimha, B., Raghavendran, C.V., Rajyalakshmi, P., Reddy, G.K., Bhargavi, M., & Naresh, P. (2022). Cyber defense in the age of artificial intelligence and machine learning for financial fraud detection application. *IJEER 10,* 87–92. https://doi.org/10.37391/ijeer.100206

Nicholls, J., Kuppa, A., & Le-Khac, N. A. (2021). Financial cybercrime: A comprehensive survey of deep learning approaches to tackle the evolving financial crime landscape. *IEEE Access, 9,* 163965–163986.

Nwankwo, C. A., & Okeke, P. O. (2021). Card fraud in Nigerian banks: Causes, consequences, and control measures. *Journal of Financial and Economic Analysis, 7*(1), 90-105.

Owolabi, K. (2020). Understanding fraud trends in Nigerian banks: The case of deposit money banks in Southeast Nigeria. *Journal of Financial Studies, 12*(4), 223-237.

Patil, A., & Suryawanshi, A. (2021). Artificial intelligence in financial fraud detection: Current applications and future trends. *Applied Intelligence, 51*(5), 1-14.

Peña, I.P., & Ortega-Castro, J.C. (2024). Implementation and evaluation of an anti-fraud prototype based on generative artificial intelligence for the Ecuadorian financial sector. *Revista de Gestão Social e Ambiental, 18*(9), 1-10, e08601 https://doi.org/10.24857/rgsa.v18n9-162

Rahman, A. (2024). IT Project Management Frameworks: Evaluating Best Practices and Methodologies for Successful IT Project Management. *Academic Journal on Artificial Intelligence, Machine Learning, Data Science and Management Information Systems, 1*(01), 57-76. https://doi.org/10.69593/ajaimldsmis.v1i01.128

Shirodkar, N., Mandrekar, P., Mandrekar, R. S., Sakhalkar, R., Kumar, K. M. C., & Aswale, S. (2020). Credit card fraud detection techniques – A survey (International Conference on Emerging Trends in Information Technology and Engineering (ic-ETITE)), 1-7. https://doi.org/10.1109/ic-etite47903.2020.112

Wei, S., & Lee, S. (2024). Financial Anti-Fraud Based on Dual-Channel Graph Attention Network. *Journal of Theoretical and Applied Electronic Commerce Research, 19*(1), 297-314. https://doi.org/10.3390/jtaer19010016